Nonlinear Optical Response of a Gold Surface in the Visible Range:

A Study by Two-Color Sum-Frequency Generation Spectroscopy.

Part II: Model for Metal Nonlinear Susceptibility

B. Busson[1], L. Dalstein[1,2,3]

(1) Laboratoire de Chimie Physique, CNRS, Univ. Paris-Sud, Université Paris-Saclay, Bâtiment 201 P2, 91405 Orsay, France

(2) Department of Chemistry, School of Chemical Science and Engineering, KTH Royal Institute of Technology, SE-100 44 Stockholm, Sweden

(3) Institute of Physics, Academia Sinica, Taipei 11529, Taiwan, Republic of China

Abstract

We present a modeling of the nonlinear optical response of a metal surface in order to account for recent experimental results from two-color Sum-Frequency Generation experiments on gold. The model allows calculating the surface and bulk contributions, and explicitly separates free and bound electron terms. Contrary to the other contributions, the perpendicular surface component is strongly model-dependent through the surface electron density profiles. We consider three electron density schemes at the surface, with free and bound electrons overlapping or spilling out of the bulk, for its calculation. The calculated SFG signals from the metal rely only on bulk quantities and do not need an explicit definition of the density profiles. In the particular case of gold, when the free electrons overlap with the bound ones or spill out of the bulk, the free electron response completely dominates through the perpendicular surface terms. When the bound electrons spill out, the situation is more balanced, still in favor of the free electrons, with lower amplitudes and different dispersion lineshapes. As for silver, the free electron contributions dominate, and the calculated slow amplitude growth from blue to red follows the experimental trends.

I. Introduction

In a previous paper[1], we have measured the experimental effective nonlinear susceptibility of a polycrystalline gold film in an infrared-visible Sum-Frequency Generation (SFG) experiment while tuning the visible wavelength over the visible range (435-705nm) (two-color Sum-Frequency Generation, 2C-SFG). Using the vibrationnally resonant response of an adsorbed thiol monolayer, we could extract the absolute phase and amplitude lineshape of gold in these conditions. In this paper, we propose to model the nonlinear response of metals, while explicitly accounting for free and bound electron contributions, as the base for a simulation of the experimental data.

Such models have been extensively develop ped for second-harmonic generation (SHG), and far less for SFG. They consist of two parts: the first one phenomenologically describes the generation process itself, determining the sources of nonlinear polarization and deriving the expressions of the generated electromagnetic waves, both at the interface and in the far field, they give birth to. The second part establishes the link between a solid state physics description of a metal bulk and surface, and the nonlinear coefficients introduced in the phenomenological part.

Phenomenological models describing the nonlinear optical response at the interface between two materials appeared soon after the discovery of nonlinear optics[2]. The formulations for the SHG nonlinear polarization produced in reflection by a surface and for the electric fields radiated in the far field were derived by several authors[3–9], and the essentials are summarized below. The nonlinear sources decompose into bulk and surface terms, even if part of the bulk response also formally contributes to an equivalent surface dipolar source, making it difficult to experimentally separate intrinsic bulk and surface contributions[6,10,11]. Over time, several refinements have improved the description of the nonlinear processes at the surface of a centrosymmetric bulk: an unambiguous definition of the quantities involved (polarization sources, nonlinear susceptibilities) has been derived[4,11]; a rigorous separation between surface and bulk contributions, each assigned to a different Green function formalism for the generation of the second harmonic electric field[5] has been established; symmetry rules have been applied to both surface and bulk contributions, reducing the number of independent susceptibility coefficients, in particular for isotropic and cubic bulks[9,12,13].

As for SFG, emphasis has early been put on its specificity in the infrared-visible configuration, namely the molecular response[14] adsorbed on a substrate. The substrate contribution is often reduced to a vibrationnally constant term, either used as a support for the interference with the molecular contribution for phase analysis[15,16], or on the contrary suppressed by pulse shaping and time-delay methods[17]. The phenomenological description of SFG of the surface and bulk from a material has therefore been given much less attention than in SHG, except in specific cases for which the bulk contribution is important like chiral liquids[18], quadrupolar SFG response[19] or for a clear separation between the bulk and surface contribution[20]. It has anyway benefitted from all the knowledge gained for the SHG case. For an isotropic system, the basic equations may be found in Ref. 2 and 21. We note that the form of the bulk contribution is much more complicated than for SHG.

For the analysis of experiments performed with fixed incoming wavelengths, or far from resonant phenomena, the phenomenological model may be sufficient provided that constant values are assigned to the various $\chi^{(2)}$ components, extracted from the data recorded as a function of an external parameter (for example the polarization states of light[10,22], angles of incidence[23], and azimuth angle[24]). For a spectroscopic study[23–25], one must take dispersion into account as the various nonlinear susceptibility tensor components will vary with the incoming wavelengths, whereas simple rules (e.g. Miller's rule) do not apply in resonant conditions. Such studies require modeling the sources of the SHG and SFG as a function of the wavelengths and material properties. For metals, the source of nonlinearity is the electrons, separated in conduction (free) and valence (bound) electrons as a consequence of the band structure. For SHG, nonlinear coefficients stem from electronic properties through the so-called Rudnick-Stern coefficients[10,25–27], which reduce the formalism to three dispersive numbers a, b and d (standing for surface perpendicular, surface parallel and bulk components, respectively)[4,26]. Most of such analyzes of the origin of $\chi^{(2)}$ devote to the free electron response only, for which the R-S coefficients were originally estimated close to unity. This is well suited for some metals (e.g. aluminum and silver[25]) but of limited use for gold in the visible range as bound electrons contribute to the optical properties of gold above 2eV through interband transitions. A simple free electron model has been shown inadequate to describe the nonlinear response of gold in the visible range[25]. Even for the free electron gas, some resonances as a function of the frequency are expected, mostly in the high energy range towards the plasma frequency of the metal[28,29]. Some authors have analyzed separately the interband contributions[9,27,30,31], but there is no consensus on whether they participate to SHG through an additional contribution to $\chi^{(2)}$ or only by modifying the linear dielectric function[27,30,32,33]. Other groups use density functional theory [28,34] or a generalized Thomas Fermi model [35] in order to circumvent the problem of *a priori* definition of the electron density profile. Taking into account at the same time both free and bound electron contributions and their interaction in nonlinear optics still remains a challenge. Using a long history in modeling first the electromagnetic fields at surfaces[36,37], then the nonlinear SHG and SFG responses of electrons in metals[7,38,39], one group proposed a way to implement the complex electronic structure of transition metals in the interband transition regime into the Rudnick-Stern coefficients[40].

In this paper, we evaluate the nonlinear responses of gold and silver in a 2C-SFG experiment in their theoretical frame[38,40] in order to compare them to the experimental data in a forthcoming step. The metal is described by its free and bound electron densities and contributions to the total dielectric function. We investigate three schemes for the nonlinear optical response, which differ by the relative positions of their electron density profiles as a function of depth and the associated coupling between free and bound contributions. We show that the free electrons dominate the total SFG response, but that the amplitude and dispersion of the effective nonlinear susceptibility of gold, and the weight of the bound electron terms, strongly depend on the profile chosen.

## II. Origin and calculation of the surface and bulk contributions to SFG

The phenomenological description is reviewed in Appendix A. In order to apply it, we must evaluate the bulk polarization, and the surface parameter $\chi_S$ after an additional integration step over z. This requires to explicitly model the electron density of the metal (i.e. its material properties) and link it to its optical response. We consider the metal as a continuous assembly of dipoles described by the classical model for the displacement of an electron, bound to its rest position by a harmonic force and a submitted to both a resisting force and the time-dependent electromagnetic field of light[38,40]. This harmonic oscillator approximation has been shown consistent with more accurate descriptions of SHG[7] and allows a calculation of the contribution to the material nonlinear response. Such a description may constitute a good approximation for a bound electron, for which a full quantum mechanical calculation would however rigorously be more appropriate[30], or a free electron when the restoring force vanishes. Solving the equations for the displacements in a Taylor expansion at second order in the electric fields and keeping only the $\omega_3$ term gives the expression of the nonlinear polarization $\mathbf{P}_B(\mathbf{r},\omega_3)$. The free electron case may also be recovered by considering the hydrodynamic equations for the velocity field of the free electron gas, while neglecting the pressure term[9,21,38]. Refined theories exist which explicitly take this extra term into account[4,39,41]. Few examples exist of a theory comprising both free and bound electron contributions with no *ad hoc* or unknown parameter included[9,30,31].

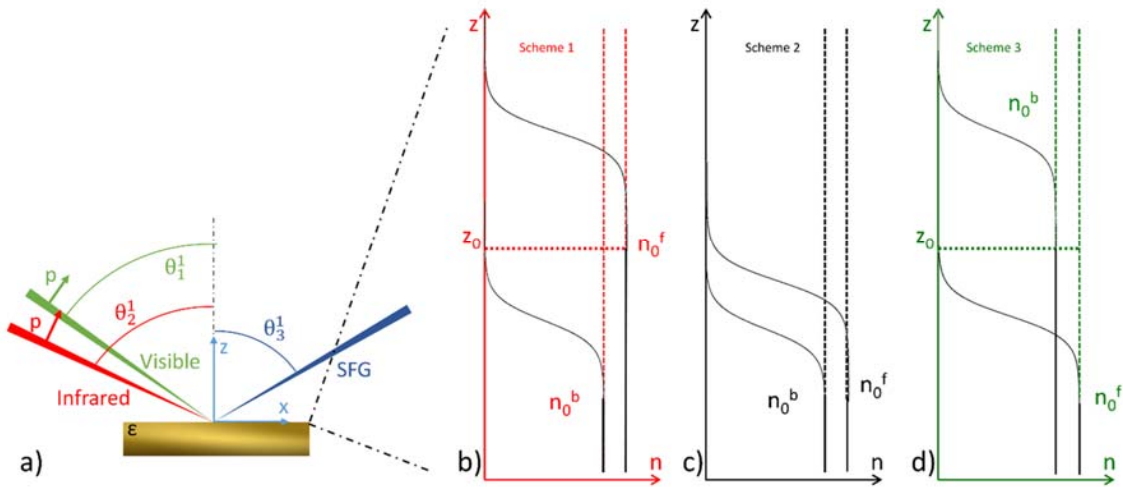

*Figure 1: (a) Sketch of the experimental configuration. (b,c,d) Schemes of the distributions of the electron densities $n^f(z)$ (resp. $n^b(z)$) of free (resp. bound) electrons as a function of depth inside the metal for schemes 1 (b), 2 (c), and 3 (d).*

The metal is described as the superimposition of two electron populations, a free electron gas (density $n_0^f$) and a bound electron density ($n_0^b$), with $n_0^f + n_0^b = n_0$. The gold interband contribution from the bound electrons shall not be neglected when considering that, for a threshold around 2eV[42], the SFG beam in our experiments is always able to excite them. For silver, the energies involved do not allow to directly excite the interband transitions, but the bound electrons still contribute to the total dielectric function. This implies that the off-resonance approximation sometimes made in the presence of interband transitions cannot apply here, at least for gold[27,30]. Starting from here, all the quantities defined in the phenomenological model have to be split between free and bound electron terms when appropriate, with the corresponding superscripts. The bulk dielectric function of the metal is therefore written as[40]:

$$\varepsilon_i \equiv \varepsilon(\omega_i) = 1 + 4\pi\alpha_i^f n_0^f + 4\pi\alpha_i^b n_0^b \quad (1)$$

where $\alpha_i^f$ (resp. $\alpha_i^b$) represent the electronic polarizabilities of the free (resp. bound) electrons at frequency $\omega_i$, and separated into the two components by $\varepsilon_i - 1 = (\varepsilon_i^f-1) + (\varepsilon_i^b-1)$, where $\varepsilon - 1$ represents the resonant part of the dielectric functions. The polarizabilities quantify the amplitude of the electronic response to the excitation by light, and the second order nonlinear response is therefore proportional to a product of two electronic polarizabilities. As the free and bound electronic polarizabilities differ, their contributions to the SFG signal must be evaluated separately, then summed up, immersed in a medium which dielectric function encompasses free and bound parts together.

The bulk polarization is obtained by plugging the electric fields, considered as plane waves with wavevectors $\mathbf{k}_1$ and $\mathbf{k}_2$ propagating inside medium 2, into the nonlinear polarization:

$$\mathbf{E}_i(\mathbf{r}) \equiv \mathbf{E}(\mathbf{r},\omega_i) = \mathbf{E}_i e^{i\mathbf{k}_i \cdot \mathbf{r}}, \; i=(1,2) \quad (2)$$

The complex wavevectors are evaluated by continuity of the component parallel to the surface and with the help of the dispersion relation of the metal $(\mathbf{k}_i)^2 = \varepsilon_i \omega_i^2/c^2$ for the perpendicular one. The field gradients in the bulk arise from the propagation of the plane waves through their phases. This leads to:

$$\mathbf{P}_B^{f/b}(\mathbf{r},\omega_3) = i e^{i(\mathbf{k}_1+\mathbf{k}_2)\cdot\mathbf{r}} \left[ D_1^{f/b} \mathbf{k}_1(\mathbf{E}_1 \cdot \mathbf{E}_2) + \Delta_1^{f/b}(\mathbf{k}_1 \cdot \mathbf{E}_2)\mathbf{E}_1 \right.$$
$$\left. + D_2^{f/b} \mathbf{k}_2(\mathbf{E}_1 \cdot \mathbf{E}_2) + \Delta_2^{f/b}(\mathbf{k}_2 \cdot \mathbf{E}_1)\mathbf{E}_2 \right] \quad (3a)$$

with $\mathbf{P}_B(\mathbf{r},\omega_3) = \mathbf{P}_B^f(\mathbf{r},\omega_3) + \mathbf{P}_B^b(\mathbf{r},\omega_3)$ \quad (3b)

where coefficients $D_i^{f/b}$ and $\Delta_i^{f/b}$ are given by equations (45) to (47) of Ref. 38 for which the electron density and dielectric functions refer respectively to free and bound electrons. We recall their expressions in Appendix B. When working with a single crystal, additional terms, which depend on the crystal orientation at the surface, should be added to account for the bulk anisotropic contribution[5,13]. We note that $D_1$ and $D_2$ have a purely dipolar origin, as well as $\Delta_2$ when the approximation $\omega_2 \ll \omega_3$ is valid. The electric field generated by this bulk polarization follows a complex expression recalled in Ref. 38, equation (A9). This equation may be tracked back through Ref. 21 to the original work of Bloembergen and Pershan[2].

The electric field gradients giving birth to the surface terms have a completely different origin. They arise in the z direction from the rapid variations of the electron densities when crossing the interface. The long wavelength approximation ensures continuity of the parallel components of the electric fields and the perpendicular ones of the displacement fields along the electron density gradients[43], and validates the dropping of the terms related to field curls in the dipolar term of the nonlinear polarization in the selvedge region[38]. In order to evaluate the surface terms, the local nonlinear polarization is calculated from equation (A1) as a function of z, while both the electron densities and the dielectric functions

$$\varepsilon_i(z) = 1 + 4\pi\alpha_i^f n^f(z) + 4\pi\alpha_i^b n^b(z) \qquad (4)$$

become z-dependent in the interface region[43,44], then integrated between external medium and bulk as usual[5,45] to calculate the surface nonlinear susceptibility components. Equation (A1) becomes:

$$\mathbf{P}_B^{f/b}(z) = n^{f/b}(z)\alpha_3^{f/b}\mathbf{E}_3 - \frac{n^{f/b}(z)}{e}\alpha_3^{f/b}(\alpha_1^{f/b}\mathbf{E}_1 \cdot \nabla\mathbf{E}_2 + \alpha_2^{f/b}\mathbf{E}_2 \cdot \nabla\mathbf{E}_1)$$
$$+ \frac{1}{2e}\alpha_1^{f/b}\alpha_2^{f/b}\nabla \cdot n^{f/b}(z)(\mathbf{E}_1\mathbf{E}_2 + \mathbf{E}_2\mathbf{E}_1) \qquad (5)$$
$$- \frac{1}{2e}\alpha_1^{f/b}\alpha_2^{f/b}\frac{\omega_2 - \omega_1}{\omega_3}\nabla \times n^{f/b}(z)\mathbf{E}_1 \times \mathbf{E}_2$$

where the first term is the depolarization part of the SFG field. Each line in equation (5) relates to the dipolar, quadrupolar and magnetic contributions, respectively, and the gradients now only relate to the z-direction. As a consequence, the dipolar (resp. magnetic) contribution does not appear in the parallel (resp. perpendicular) component of the nonlinear polarization.

The origin of the dipolar term is different from the quadrupolar and magnetic contributions. For the former, direct coupling of one incoming electric field and one electric field gradient generates a polarization at the SFG frequency proportional to the SFG polarizability $\alpha_3$. For the latter, the coupling of the incoming electric fields generates local quadrupoles and magnetic dipoles, which gradients take part of the total polarization. This accounts for their proportionalities to polarizabilities at the incoming frequencies only. This has major consequences in the IR-visible SFG case applied to metals like gold and silver, which may be qualitatively described by $\omega_2 << (\omega_1, \omega_3)$ and $|\alpha_2^b| << |\alpha_2^f|$. As will be checked below, we therefore expect that the magnetic and quadrupolar contributions to the bound electron polarization should be negligible compared to free electron terms. This implies that, for bound electrons, the total parallel surface contribution is also small, whereas the bulk and perpendicular terms are from dipolar origin only. As for the free electron contributions, it has already been noticed by Petukhov[21] that the $\Delta_1$ and $\Delta_2$ terms almost vanish for a free electron gas, leaving $D_1 \approx D_2$ as leading terms. As a consequence, it appears that the full bulk contribution (i.e. for free and bound electrons) is purely dipolar under reasonable approximations. This contributes to simplify the modelling of the IR-visible SFG bulk processes in metals.

The free (resp. bound) electron component of the polarization parallel to the surface $P_{B,x}^f(z)$ (resp. $P_{B,x}^b(z)$) becomes:

$$P_{B,x}^f(z) = \frac{\alpha_1^f \alpha_2^f}{2e}\left[\left(1 - \frac{\omega_2 - \omega_1}{\omega_3}\right)\frac{\partial}{\partial z}\frac{n^f(z)}{\varepsilon_2(z)}\varepsilon_2 E_{1x}E_{2z} + \left(1 + \frac{\omega_2 - \omega_1}{\omega_3}\right)\frac{\partial}{\partial z}\frac{n^f(z)}{\varepsilon_1(z)}\varepsilon_1 E_{1z}E_{2x}\right] \qquad (6)$$

The difference between $\omega_1$ and $\omega_2$ induces a partial amplitude transfer from xzx to xxz terms, as a consequence of the interference between magnetic and quadrupolar contributions. Integration according to equation (A2) is straightforward, and the xxz and xzx components therefore do not depend on the electron density profiles across the interface[40]. Integral leads to the following expressions for the xxz and xzx components of $\chi_S$ :

$$\chi_{S,xxz}^{f/b} = \frac{1}{2n_0^{f/b}e}\frac{\varepsilon_1^{f/b}-1}{4\pi}\frac{\varepsilon_2^{f/b}-1}{4\pi}\frac{2\omega_1}{\omega_3}b^{f/b}(\omega_1,\omega_2) \quad (7a)$$

$$\chi_{S,xzx}^{f/b} = \frac{1}{2n_0^{f/b}e}\frac{\varepsilon_1^{f/b}-1}{4\pi}\frac{\varepsilon_2^{f/b}-1}{4\pi}\frac{2\omega_2}{\omega_3}b^{f/b}(\omega_2,\omega_1) \quad (7b)$$

$$\chi_{S,xxz} = \chi_{S,xxz}^{f} + \chi_{S,xxz}^{b} = \frac{1}{2n_0 e}\frac{\varepsilon_1-1}{4\pi}\frac{\varepsilon_2-1}{4\pi}\frac{2\omega_1}{\omega_3}b(\omega_1,\omega_2) \quad (7c)$$

$$\chi_{S,xzx} = \chi_{S,xzx}^{f} + \chi_{S,xzx}^{b} = \frac{1}{2n_0 e}\frac{\varepsilon_1-1}{4\pi}\frac{\varepsilon_2-1}{4\pi}\frac{2\omega_2}{\omega_3}b(\omega_2,\omega_1) \quad (7d)$$

with the simple values $b^{f/b}(\omega_1,\omega_2)= b^{f/b}(\omega_2, \omega_1) = -1$. The last two equations define the overall $\chi_S$ components by adding the free and bound electron contributions. Care must be taken as global parameters $b(\omega_1,\omega_2)$ and $b(\omega_2,\omega_1)$ now differ from -1.

On the contrary, because of the dipolar contribution, integration of the perpendicular component (zzz) has to be explicitly performed. No term in the selvedge nonlinear polarization arises from two electric fields parallel to the surface, leading as usual to a vanishing zxx contribution to $\chi_S$ [21,38]. For the zzz term, the z-components of the selvedge nonlinear polarizations induced by the free (resp. bound) electrons $P^f_{B,z}(z)$ (resp. $P^b_{B,z}(z)$), evaluated inside the metal, are

$$P^f_{B,z}(z) = \frac{\varepsilon_1\varepsilon_2\varepsilon_3}{e\varepsilon_3(z)}\left[-n^f(z)\alpha_3^f\left(\frac{\alpha_1^f}{\varepsilon_1(z)}\frac{\partial}{\partial z}\frac{1}{\varepsilon_2(z)}+\frac{\alpha_2^f}{\varepsilon_2(z)}\frac{\partial}{\partial z}\frac{1}{\varepsilon_1(z)}\right)\right.$$
$$\left.+\alpha_1^f\alpha_2^f\frac{\partial}{\partial z}\left(n^f(z)\frac{1}{\varepsilon_1(z)}\frac{1}{\varepsilon_2(z)}\right)\right]E_{1z}E_{2z} \quad (8)$$

where the quantities not explicitly depending on z are the bulk ones, and one has to integrate them as in equation (A2). Following equation (4), the integrand becomes a function of $n^f(z)$ and $n^b(z)$ only, which are supposed continuous in the following. In order to actually calculate these zzz terms, we therefore must make some assumptions on these quantities.

We consider the three extreme cases for the electron distributions: two split electron distributions (Figure 1(b),(d)) and an overlapping electron model (Figure 1(c)). In the first scheme (s1, Figure 1(b)), the free electrons spill out from the bulk whereas bound electrons don't, so their respective electron density gradients do not overlap as a function of depth. It is known that, for a free metallic surface, the free electrons may spill out of the bulk (over a depth far below one nanometer)[33,43], thus enhancing their contribution to the surface terms. We thus consider that the free electron density as a function of depth varies from zero to its bulk value $n_0^f$ (reached for $z=z_0$) while $n^b(z)=0$, then remains constant for $z<z_0$ while the interband electron density raises from zero to its bulk value $n_0^b$ (Figure 1(b)). For a complete analysis, we may consider two additional configurations: in scheme 2 (s2, Figure 1(c)), the free and bound electron distributions overlap as a function of depth, whereas in scheme 3 (s3, Figure 1(d)), the bound electrons spill out more than the free electrons symmetrically to scheme 1. As stated above, the model chosen for the electron density profiles has no influence on the bulk and parallel surface terms. Consequently, equations (3) and (7) remain unchanged for schemes 1, 2 and 3 and we concentrate in the evaluation of the perpendicular zzz term.

Integration of the z-component of the nonlinear polarization in scheme 1 leads to a sum of three contributions: a pure free electron ($a^f$) one arising from the surface above $z_0$; a free electron contribution induced by the variation in the bound electron density across the interface ($a^{bf}$); a pure bound electron contribution ($a^b$) screened by the bulk free electron density:

$$\chi^{s1,f}_{S,zzz} = \frac{1}{2n_0^f e} \frac{\varepsilon_1^f - 1}{4\pi \varepsilon_1^f} \frac{\varepsilon_2^f - 1}{4\pi \varepsilon_2^f} \varepsilon_1 \varepsilon_2 \varepsilon_3 \left[ a_{s1}^f(\omega_1, \omega_2) + a_{s1}^{bf}(\omega_1, \omega_2) \right] \quad (9a)$$

$$\chi^{s1,b}_{S,zzz} = \frac{1}{2n_0^b e} \frac{\varepsilon_1^b - 1}{4\pi \varepsilon_1^b} \frac{\varepsilon_2^b - 1}{4\pi \varepsilon_2^b} \varepsilon_1 \varepsilon_2 \varepsilon_3 a_{s1}^b(\omega_1, \omega_2) \quad (9b)$$

and $\chi^{s1}_{S,zzz} = \chi^{s1,f}_{S,zzz} + \chi^{s1,b}_{S,zzz} = \frac{1}{2n_0 e} \frac{\varepsilon_1 - 1}{4\pi} \frac{\varepsilon_2 - 1}{4\pi} \varepsilon_3 a_{s1}(\omega_1, \omega_2) \quad (9c)$

defines the global parameter $a_{s1}$, with

$$a_{s1}^f(\omega_1, \omega_2) = -2 \left[ 1 + \frac{(1-\varepsilon_3^f)\varepsilon_1^f \varepsilon_2^f}{(\varepsilon_1^f - \varepsilon_2^f)(\varepsilon_2^f - \varepsilon_3^f)(\varepsilon_3^f - \varepsilon_1^f)} \sum_{\substack{\{i,j,k\}= \\ \text{c.p.of} \\ \{1,2,3\}}} \varepsilon_i^f \ln\left(\frac{\varepsilon_j^f}{\varepsilon_k^f}\right) \right] \quad (10a)$$

where c.p. denotes circular permutation of {1,2,3},

$$a_{s1}^{bf}(\omega_1, \omega_2) = -\frac{2}{\varepsilon_3^f} \left[ \frac{\varepsilon_1^f \varepsilon_2^f \varepsilon_3^f}{\varepsilon_1 \varepsilon_2 \varepsilon_3} - 1 + \sum_{i=1}^{3} \frac{\varepsilon_3^f - 1}{\varepsilon_i^f - 1} \frac{\varepsilon_i^b - 1}{\varepsilon_i^f} I_0\left(\frac{\varepsilon_i^b - 1}{\varepsilon_i^f}\right) \right] \quad (10b)$$

$$a_{s1}^b(\omega_1, \omega_2) = -\frac{2\varepsilon_1^b \varepsilon_2^b}{\varepsilon_1^f \varepsilon_2^f \varepsilon_3^f} \left[ \frac{\varepsilon_1^f \varepsilon_2^f \varepsilon_3^f}{\varepsilon_1 \varepsilon_2 \varepsilon_3} + \sum_{i=1}^{3} \frac{\varepsilon_3^b - 1}{\varepsilon_i^f} I_1\left(\frac{\varepsilon_i^b - 1}{\varepsilon_i^f}\right) \right] \quad (10c)$$

The technical details of the integration steps, along with definitions and properties of $I_0$ and $I_1$ integrals can be found in the Appendix C. The result is identical to Ref. 40, while we correct here some original confusion in the attribution of each term. For scheme 3, the results are identical provided that superscripts b and f are exchanged in equations (9) and (10).

In scheme 2, the integration is performed in the same way as above, leading to the new values for the free electron and interband contributions to parameter $a_{s2}$.

$$\chi^{s2,f}_{S,zzz} = \frac{1}{2n_0^f e} \frac{\varepsilon_1^f - 1}{4\pi \varepsilon_1^f} \frac{\varepsilon_2^f - 1}{4\pi \varepsilon_2^f} \varepsilon_1 \varepsilon_2 \varepsilon_3 \left[ a_{s2}^f(\omega_1, \omega_2) \right] \quad (11a)$$

$$a_{s2}^{f}(\omega_1, \omega_2) = -2\left[\frac{\varepsilon_1^f \varepsilon_2^f \varepsilon_3^f}{\varepsilon_1 \varepsilon_2 \varepsilon_3} + \frac{(1-\varepsilon_3^f)\varepsilon_1^f \varepsilon_2^f}{(\varepsilon_1-\varepsilon_2)(\varepsilon_2-\varepsilon_3)(\varepsilon_3-\varepsilon_1)} \sum_{\substack{\{i,j,k\}= \\ c.p.of \\ \{1,2,3\}}} \varepsilon_i^f \ln\left(\frac{\varepsilon_j^f}{\varepsilon_k^f}\right)\right.$$
$$\left. + (\varepsilon_3^f - 1)\varepsilon_1^f \varepsilon_2^f \left(\sum_{i=1}^{3} \frac{\varepsilon_i^b - 1}{\varepsilon_i^f - 1} I_1(\varepsilon_i - 1)\right)\right] \qquad (11b)$$

For the interband contribution, one may simply exchange $\varepsilon_i^f$ and $\varepsilon_i^b$ in equations (11). We also may check that formula (11b) turns down to (10a) when the bound electron contribution vanishes ($\varepsilon_i^b = 1$ and $\varepsilon_i = \varepsilon_i^f$).

We immediately see that the zzz contribution (as well as the parallel one for obvious reasons) depends neither on the spatial evolution of the electron density distributions (i.e. $n^{f/b}(z)$ are arbitrary continuous functions) nor on any intrinsic length characteristic of the interface, related for example to the thickness of the interface layer or to the spatial extent of the electron density gradients. For that respect, the present model allows by its simplicity to avoid the delicate definition of the density profiles, which has been shown to raise severe issues in the hydrodynamic model [35]. On the contrary, it only depends on the scale-free relative positions of the electron densities, and of course on their bulk values. This means that, in this modeling, the magnitudes of the gradients (of the electric fields and electron densities) are compensated by their spatial extents after integration into the surface nonlinear polarization. However, as we will see below, integration leads to huge differences between the various schemes, in particular as a consequence of self- and cross-screening between free and bound contributions.

In order to illustrate the results of the modelling of the nonlinear response of metals, we briefly present the results for the simple case of silver, then we focus on the more complex case of gold, for which a detailed analysis will be performed in the third part of this article.

### III. The nonlinear SFG response of silver

The dielectric function of silver is tabulated from Yang *et al.*[46] (sample C, corrected for Kramers-Kronig consistency). In order to separate the free and bound electron parts, we use their fits of the free electron behavior by a Drude model in the near infrared range, with parameters $\omega_P$=8.90eV and $\Gamma$=36.57meV.

$$\varepsilon_i^f = 1 - \frac{\omega_P^2}{\omega_i(\omega_i + i\Gamma)} \qquad (12)$$

In the visible range, the bound electron dielectric function follows: $\varepsilon_i^b = 1+\varepsilon_i - \varepsilon_i^f$ for i=(1,3). It encompasses a contribution from $\varepsilon_\infty$, which we suggest to correct to 4 instead of the tabulated value of 5, in order to comply with the original data of Ref.    . For the infrared beam, the remaining bound electron contribution is supposed real but unknown, we therefore performed our simulations with $\varepsilon_2^b$=2 [40].

The various contributions to the effective $\chi^{(2)}$ for silver are displayed on Figure 2. As may be anticipated, the effect of the interband terms is rather weak, especially at low energy, except marginally for scheme 3. The free electron contribution dominates, for scheme 1 and 2 through the zzz term, and for scheme 3 through the surface parallel and bulk terms. All contributions share a common phase within a 30° range, except for a sign change for the bound and cross terms in scheme 3.

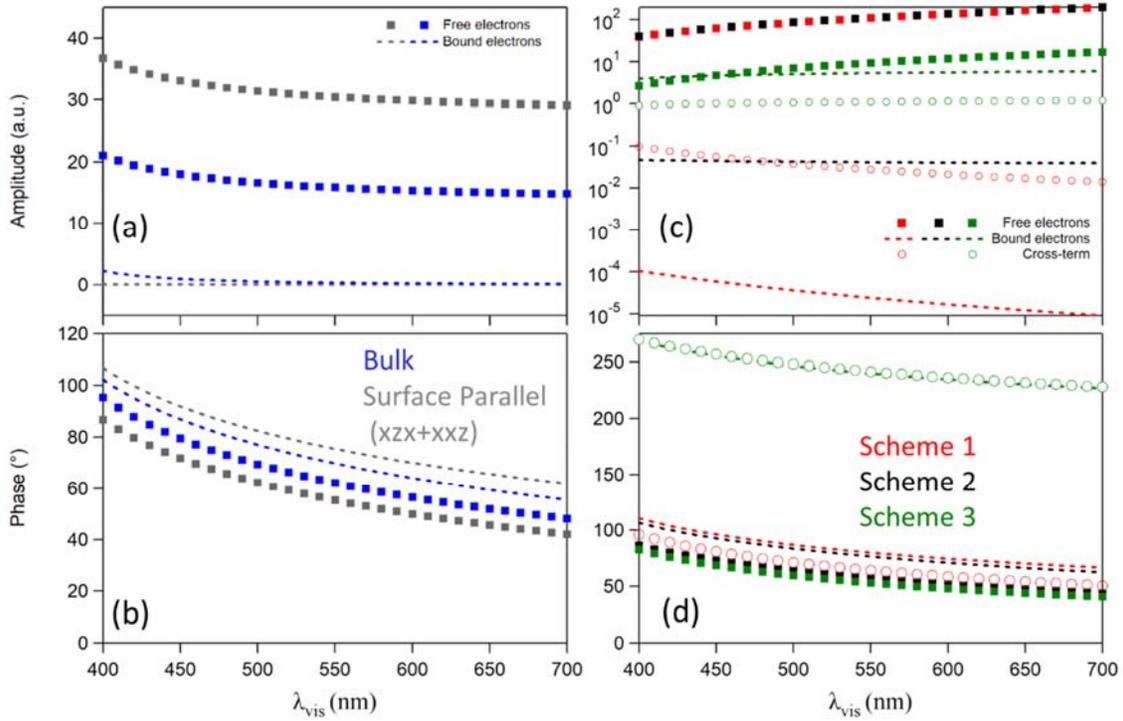

Figure 2: Amplitudes (a,c) and phases (b,d) of the various components of surface parallel (xzx+xxz), bulk (a,b) and surface perpendicular zzz (c,d) contributions to the effective nonlinear susceptibilities of silver. On the left, red color stands for surface parallel term and blue for bulk. Squares: free electrons, dashed lines: bound electrons. On the right, red, black and green stand for schemes 1, 2 and 3, respectively. Squares: free electrons, dashed lines: bound electrons, open circles: cross term (bf). Infrared wavenumber is fixed at 2900cm$^{-1}$ and $\varepsilon_2^b$=2.

Figure 3 shows the total effective nonlinear susceptibility of silver in the visible range, for two infrared wavenumbers: 2900cm$^{-1}$ (representative of studies in the CH stretch region) and 1000cm$^{-1}$ (for the phenyl ring collective modes), under the hypothesis $n_0^b = n_0^f$. Schemes 1 and 2 compare very well, showing the expected free electron behavior, with a regular increase in amplitude from blue to red. Scheme 3 differs as a consequence of the more pronounced interference between the three free electron terms, which compensate each other to exhibit a rather flat total contribution. The relative amplitude between schemes 1 and 2 on one side and 3 on the other side varies from 2 to 5 from blue to red. The phases for the three schemes are the same. These curves do not change for a ratio $R_{bf}=n_0^f/n_0^b$ varying from 0.1 to 1 (dominant free electron character). For the low energy infrared case, the amplitudes increase (for schemes 1 and 2) by up to 20 percent as a consequence of a higher infrared free electron polarizability, whereas the phases decrease by a low 8 to 13°.

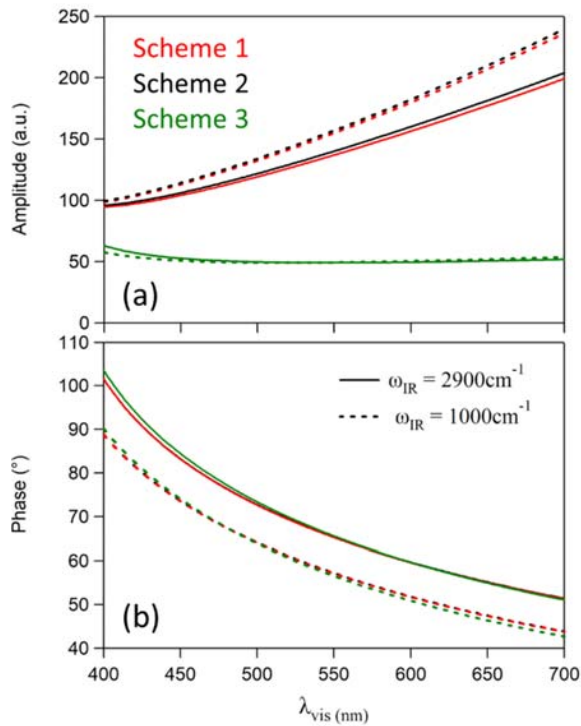

*Figure 3: Total effective susceptibility of silver in amplitude (a) and in phase (b) for schemes 1, 2 and 3. Red, black and green stand for schemes 1, 2 and 3, respectively. Infrared wavenumber is fixed at 2900cm$^{-1}$ (plain lines) or 1000cm$^{-1}$ (dashed lines). Constant numerical factors are dropped, $\varepsilon_2^b=2$. and $n_0^f=n_0^b=1$.*

Comparing these curves to experimental data remains difficult because recording the absolute SFG response of a silver bare surface requires an accurate absolute calibration of the SFG spectrometer [25] and does not provide with the phase parameter. To our knowledge, there is no such study available in the literature. After adsorption of a molecular monolayer, data analysis allows to extract a phase parameter of the silver response and an amplitude ratio at a fixed or tunable visible wavelength. As for the amplitudes of the silver SFG signals, only a few SFG experiments tune the visible wavelength on such systems [47–51] and even fewer provide with useful parameters [47,49]. Using the same data analysis as in Ref. 1, i.e. by comparing molecular and metal SFG responses, the published data which explicitly quantify the amplitude of the silver contribution are then compatible with the present model under schemes 1 and 2 (i.e. with a dominant free electron zzz response), with a silver SFG contribution slowly growing in amplitude from blue to red as compared to the molecular one.

However, comparison to the present results remains difficult in a straightforward manner because the behavior shown in Figure 3 is then modulated by the molecular SFG response [1]. The latter strongly depends on the nature of the molecule adsorbed at the silver surface: its symmetries dictates on the nonvanishing molecular hyperpolarizabilities, and its orientation (tilt and azimuth angles) impact on the relationships between molecular hyperpolarizability and nonlinear susceptibility contributions [52]. In addition, these are modulated by their own Fresnel factors, which introduce dispersion in the visible range both in amplitude and in phase, and by wavelength-dependent Raman polarizabilities, which may become resonant on the high energy side [50,51]. Finally, care must be taken that i) the silver surfaces under study are often chosen as single crystals (for which additional terms for the nonlinear susceptibility are required); ii) the results may vary between co-propagating and counter-propagating SFG geometries [53,54], and iii) several studies are performed under electrochemical control, in which case the electrochemical potential modulates the amplitude and phase of the silver signals, in particular the free electron part [25,55,56]. Experimentally, the silver SFG response remains weak (contrary to gold), although experimental evidence may vary on this point [48,57,58]. Consequently, the uncertainties on the amplitude, and even more on the phase, extracted from curve fitting of the silver response become rather big (considering that the interference pattern between molecular and weak metal nonlinear contributions is less pronounced). It has also been shown, using the example of a thiophenol monolayer on silver [57], that curve fitting may not lead to unique set of parameters [59]. We can indeed check in the literature that the fitted silver phase varies over rather wide ranges [48,50,53,57,60–62] but, without a full molecular analysis predicting the phase of the vibrational peaks [1], this parameter alone is not a useful probe of the present theory. In addition, many experiments involve molecules which are doubly resonant, or either electronically [63] or vibrationally [64] modified by adsorption on metal, which dismisses them as reliable molecular probes. There definitely lacks a dedicated study of an adapted molecule (e.g. long chain alkanethiol) adsorbed on polycrystalline silver, for which SFG spectra would be recorded and fully analyzed at enough wavelengths covering the visible and near infrared ranges.

## IV. The nonlinear SFG response of gold

We now focus on the more interesting case of gold since there is a strong interference between free and bound electron contributions in the visible range. The dielectric function is tabulated from Olmon *et al.*[65] (template-stripped sample). In order to separate the free and bound electron parts, we fit the free electron behavior by a Drude model in the near infrared range and find parameters $\omega_P$=8.80eV and $\Gamma$=49.4meV. The bound electron dielectric function is evaluated as for silver, with $\varepsilon_2^b$=2.

For the three schemes, the bound electron contribution is considered as a whole, even if the valence electrons fall into two families: two electrons in the band closest to the Fermi level (upper d-band) experience interband transitions in the visible and near-UV, while the eight other electrons lie in bands with lower energies[66] (lower d-bands). We could therefore split the bound electron contributions in two parts by considering the upper and lower d-bands as two distinct and interacting dipolium contributions. A refined model explicitly separating both contributions is currently under investigation. At this stage, for the simulations below, we follow the choice of Ref. 40 and use the arbitrary values $n_0^b = n_0^f$ in order to illustrate the balance between free and bound electron terms.

### A. Fresnel factors

The evolution of the amplitudes and phases of the Fresnel factors for zzz, xxz, xzx and bulk terms is presented on Figure 4. The absolute value for zzz is around a hundred times smaller than for xzx, and ten times smaller than for xxz. This stems from the fact that the ratio $F_x(\omega_i)/F_z(\omega_i)$ is roughly proportional to the index of refraction of gold ($n_i$). A second consequence of this is seen on the phase curves: at the low energy end of the visible spectrum, the influence of interband transitions becomes smaller and the refractive index of gold is roughly imaginary for the three beams. All Fresnel factors share at this point a common phase, except for a sign difference between $F_{xxz}$ and $F_{xzx}$ on one side, and $F_{zzz}$ and $F_{bulk}$ on the other side. From the low value of $F_{zzz}$, one may thus be tempted to neglect the contribution from the a term in the following. However, it has been shown by SHG that the weakness of the Fresnel contribution to zzz may be counterbalanced by the high value of the a source term itself[10,11], thus producing perpendicular and parallel contributions comparable in intensity[24]. The situation is analogous in SFG, although Fresnel factors for xxz and xzx additionally differ as a consequence of the intense screening of the infrared field inside gold through its high dielectric function.

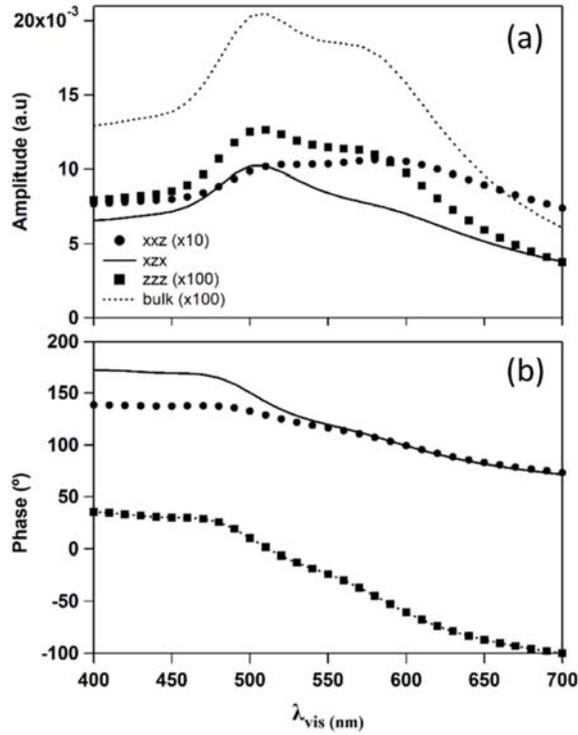

*Figure 4: Amplitudes (a) and phases (b) of the xxz (dots), xzx (plain lines), zzz (squares) and bulk (dashed lines) Fresnel factors of gold versus the visible wavelength ($\lambda_{vis}$). Infrared wavenumber is fixed at 2900cm$^{-1}$. Phases of the bulk and zzz factors are equal. Several amplitudes are multiplied for clarity by a factor 10 (xxz) or 100 (zzz, bulk).*

It is interesting to note that the amplitudes, especially for zzz, show two maxima in the middle of the visible range, related to the excitation of the interband transitions of gold by the visible and SFG beams, respectively. More precisely, considering the formulas (A6) for x and z Fresnel factors, and taking into account the fact that $\cos(\theta_i^2)$ is a complex number close to 1 with a small imaginary part, x and z contributions are maximum when $1+n_i \cos\theta_i^1$ (resp. $n_i (1+n_i \cos\theta_i^1)$) is roughly minimum, that is when the modulus of the index of refraction (or the dielectric function) is minimum. This happens around $\omega_i$=500nm, the effect being more peaked for z because of the additional $n_i$ factor. This has already been noticed by Ref. 28 for SHG, but in our case this leads to a double peak corresponding to the two beams involved in the visible range. The behavior of the zzz contribution alone may account for the existence of the experimental maximum seen in Ref. 1. At this stage, taking into account the Fresnel effects, which was not done in Ref. 67, could cast some doubt on the influence of surface electronic states as postulated therein, and attribute the experimental maxima to a simple linear optical effect. However, only the full SFG response in the far field (i.e. the effective nonlinear susceptibilities) is measured, and we need to evaluate both the intrinsic $\chi^{(2)}$ amplitudes and their Fresnel screening factors. In addition, the phase parameter gives an additional track to estimate the accuracy of the modeling.

B. Surface and bulk contributions

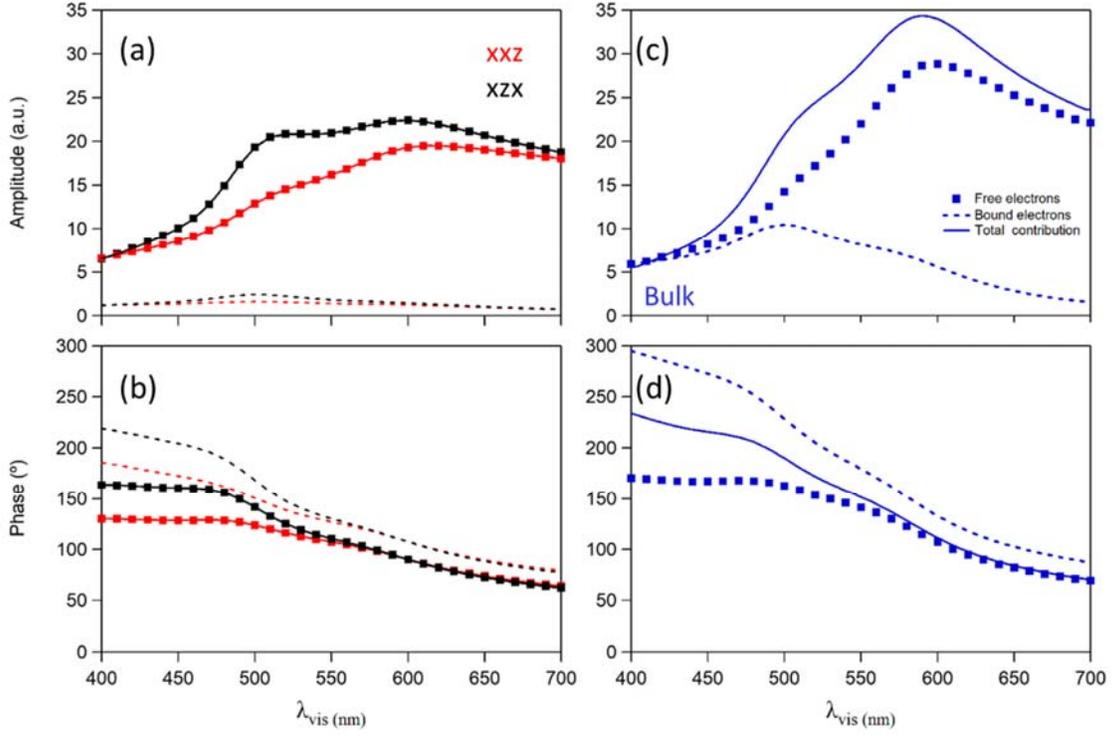

*Figure 5: Amplitudes (a,c) and phases (b,d) of the various components of xxz, xzx (a,b) and bulk (c,d) contributions to the effective nonlinear susceptibilities of gold. Red color stands for xxz term, black for xzx and blue for bulk. Squares: free electrons, dashed lines: bound electrons, plain lines: total contributions. Infrared wavenumber is fixed at 2900cm$^{-1}$ and $\varepsilon_2^b$=2. Amplitude for the interband part of xxz and xzx contributions is multiplied by 100 for clarity. Constant numerical factors are dropped and $n_0^f = n_0^b = 1$ is used to calculate the full contribution.*

The surface xxz and xzx, as well as the bulk contribution, do not depend on the scheme. Figure 5 displays the dispersion of the corresponding effective second order susceptibilities in amplitude and phase for $\varepsilon_2^b$=2. Components xxz and xzx roughly share the same amplitudes and, as expected from the above analysis, they solely originate from the free electrons. As for the bulk, free and bound electrons equally contribute at high energy, whereas free electrons dominate at the other end of the spectrum. Nevertheless, the bulk free electron contribution (growth from blue to red) is strongly influenced by the interband part of the dielectric function, as shown by the presence of two broad maxima corresponding roughly to the excitation of the interband transitions by the visible and SFG photons, with a prominent peak around 600nm. The same behavior is visible to a lower extent for surface parallel terms. The bulk interband contribution closely follows the interband dielectric function, with a profile showing a decrease towards higher wavelengths and a maximum in the 500nm region. This confirms the analysis of Ref. 30, which have shown that the bound electron contribution to the bulk SHG response is negligible except when the incoming photon energy matches that of the interband transition, as it happens in Figure 5 (c). In addition, they remark that, apart from this resonance case, their effect appears only through their modification of the dielectric function (i.e. in the free electron contribution). All phases share a close to common value at 700nm (between 62 and 90°) and a similar raise from red to blue, with a slope increasing from the xxz case (phase span around 70°) to the bulk bound electron terms (phase span around 210°).

The free electron terms do not depend on the value chosen for $\varepsilon_2^b$, whereas the interband contribution to xxz and xzx is negligible. In addition, it can be checked that the interband contribution to the bulk term does not depend either on $\varepsilon_2^b$. The leading terms for the bulk interband contribution ($D_2$ and $\Delta_2$) arise from the strong gradients of the infrared electric field and do not depend on the infrared bound electron polarizability. As a whole, the full surface parallel and bulk terms are therefore independent of the choice of the infrared dielectric function for bound electrons.

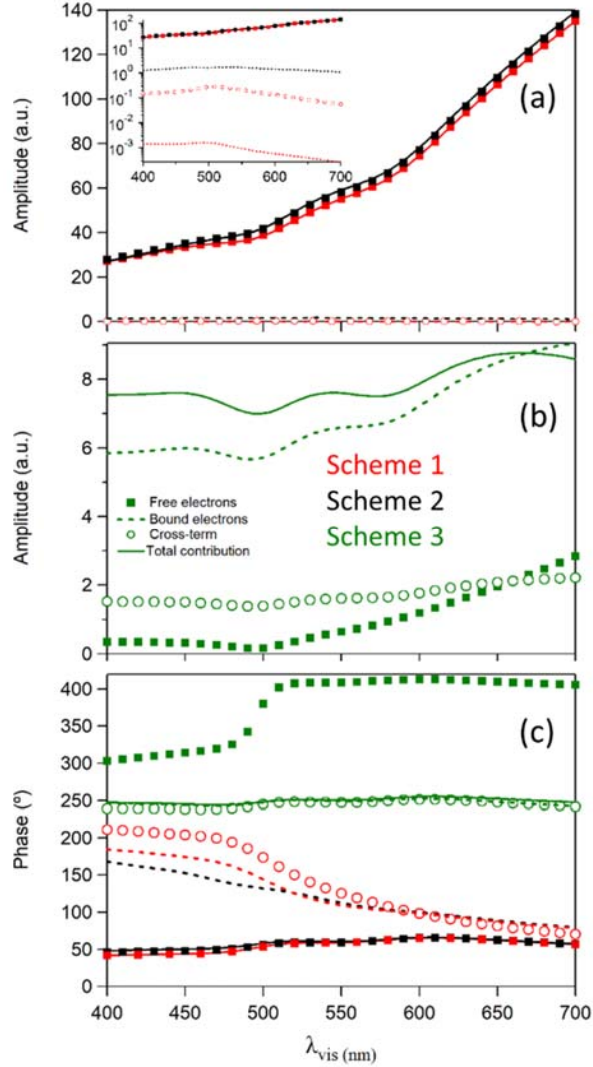

*Figure 6: Lineshapes of the amplitudes (a,b) and phases (c) of the various contributions to the effective $\chi_{zzz}$ of gold according to schemes 1 (red) and 2 (black) in (a,c) and 3 (green) in (b,c). Squares: free electrons, dashed lines: bound electrons, open circles: cross-terms (bf), plain lines: total contributions. In the inset of panel (a), a logarithmic scale allows to visualize the smallest contributions. Infrared wavenumber is fixed at 2900cm$^{-1}$ and $\varepsilon_2^b=2$. Constant numerical factors are dropped and $n_0^f=n_0^b=1$ is used to calculate the full contribution.*

We now turn to the zzz contribution to the surface response, the only model-dependent one. Figure 6 shows the dispersion in amplitude and phase of the various contributions to effective nonlinear susceptibilities in the three schemes. The behaviors of the three models are dramatically different: schemes 1 and 2 show no substantial difference, contrary to scheme 3. In schemes 1 and 2, the inner contribution involving bound electrons ($\chi^{bf}$ and $\chi^b$) is negligible (by at least a factor $10^2$ and $10^4$, respectively) as compared to $\chi^f$. This is due to the huge difference in field gradients in the infrared between free and bound electrons, in favor of the free electron response. The zzz part of the surface term therefore turns down to a nearly pure free electron response, with a characteristic increase towards the low energy region comparable to a Drude dielectric function. On the contrary, in scheme 3, the high infrared field gradients of the free electrons are screened by the bound electron background. The total response is more of the bound electron type, which accounts for the smaller overall amplitude due to its smaller field gradients. In addition, the bound electron contribution is rather flat, as is the modulus of the bound dielectric function. The phases of the dominant terms (thus also of the total contributions) stay rather constant as a function of the wavelength for the three schemes, leading to schemes 1,2 on one side and 3 on the other being almost opposite in sign. We again acknowledge the analogy with the dielectric functions, which also show a phase difference around 180° between free and bound electrons.

As for the phases of all terms in Figures 5 and 6, their evolutions are either rather flat or experience a change in their values around 500nm and/or 600nm. This corresponds to the wavelengths for which the Fresnel contributions show a maximum (and the total dielectric function a dip) as explained above, and also to the turning point where the free electron dielectric function becomes greater than the interband one. Finally, as was noticed for the Fresnel factors, all surface and bulk terms share an almost universal phase at 700nm, except for a sign difference for the zzz contribution in scheme 3. We therefore expect the phase to remain constant for the total effective susceptibility of gold at this wavelength.

Figure 6 therefore evidences a dramatic evolution from scheme 1 to scheme 3, with an effective total zzz term evolving from a pure free electron to a nearly pure bound electron contribution (with a lower amplitude). In other words, the (unscreened) external electron density mainly drives the overall behavior, with a natural higher amplitude for the free electron part. In addition, both amplitudes lineshapes and relative phases of the total contributions follow the trends of the dielectric function of their leading terms.

Tuning parameter $\varepsilon_2^b$ does not influence much the zzz contribution under schemes 1 and 2. As only the terms involving the interband polarizabilities depend on it, their weaknesses compared to the free electron contributions makes the effect of this parameter negligible. With a closer look, it may be checked that $\chi^{bf}$ (scheme 1) and $\chi^b$ (scheme 2) do not vary with $\varepsilon_2^b$. Only $\chi^b$ in scheme 1 is influenced by this parameter, but its amplitude is hundred times less than that of $\chi^{bf}$. We may therefore expect that $\varepsilon_2^b$ has no influence at all on the total nonlinear susceptibilities in schemes 1 and 2. As for scheme 3, the $\chi^b$ term increases whereas the $\chi^{bf}$ term decreases with $\varepsilon_2^b$, with very close phases. As a consequence, their sum slowly grows but in limited amounts.

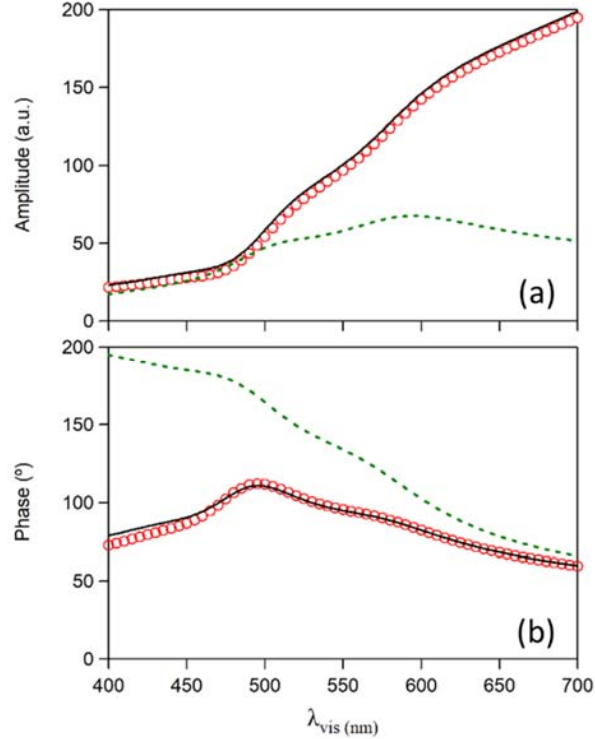

*Figure 7: Total effective susceptibility of gold in amplitude (a) and in phase (b) for schemes 1, 2 and 3. Open circles: scheme 1, plain line: scheme 2, dashed line: scheme 3. Infrared wavenumber is fixed at 2900cm$^{-1}$ and $\varepsilon_2^b=2$. Constant numerical factors are dropped and $n_0^f=n_0^b=1$.*

Comparing Figure 5 and 6, in schemes 1 and 2 the zzz absolute amplitudes prevail over the surface parallel and bulk contributions, whereas in scheme 3 the latter will dominate the full response, except at high energy. The total effective susceptibility of gold for the three schemes appears on Figure 7 under the hypothesis of equal contributions from free and bound electrons. The amplitude lineshapes in schemes 1 and 2 look very similar indeed to the zzz contribution (i.e. of pure free electron type), but the overall amplitude is higher and the resonances around 500 and 600nm change from dips to humps, showing that the effect of surface parallel and bulk contributions is small but not negligible. This is confirmed by the phase curves in Figure 7(b), which show a raise in the average phase with distinguishable resonant structures. As a whole, for these schemes, the SFG response of gold is essentially of free electron origin, except a small contribution from the bulk term. As for scheme 3, as expected the total effective susceptibility follows the trends of the surface parallel and bulk terms, with a clear effect of the two resonances on the amplitude, whereas the phase exhibits the same raise from red to blue, with intermediate values between xxz/xzx and bulk ones. Again, the total effective susceptibility is mainly of the free electron type, but the influence of the bound electron contributions from zzz and bulk terms should this time not be neglected. In addition, it may be enhanced depending on the actual value of $n_0^f/n_0^b$, which will be done in a future step.

## V. Discussion

These results show that, for schemes 1 and 2 with $n_0^f=n_0^b$, the nonlinear susceptibility of gold originates almost completely in the free electrons, as is often assumed[32]. The interband contribution may however not be omitted in the frame of scheme 3. For silver, the three schemes favor the free electron response, even if the dispersion curves differ between them. For both metals, the influence of bound electrons mostly shows up through the dielectric functions. The three schemes considered here represent extreme and schematic configurations for which the zzz component is analytically calculable without knowledge of the actual functions describing the electron density profiles. Of course, the actual profiles may be more accurately described by an intermediate configuration between two of these extreme schemes, leading to an intermediate behavior of the simulated zzz contribution (Figure 6) and total effective susceptibility (Figure 7). We plan to refine our simulations in this way in the future. However, it is not possible to consider a model for which the screening of the bound electron terms would be lower than for scheme 3.

It is interesting to compare the absolute SFG intensities radiated by a silver (Figure 3) and a gold (Figure 7) surface. The present simulations predict that their effective susceptibilities, for example for experiments with a 532nm visible laser, should compare, and, as they share common bulk free electron and valence d electron densities, so should the experimental intensities. We know that this is not the case, and this may be due to an underestimation of the actual contribution of the bound electrons of gold. In addition, as will be studied in details in the next part of this article, the excitation of surface states [67] may lead to a resonant enhancement of the surface gold response, accounting for the difference between gold and silver.

We use the bulk polarizabilities all over the calculations, including in the selvedge. It may be considered that, in this region, the polarizabilities differ from the bulk as a consequence of the change in the atomic coordination in the topmost layer. In particular, we could expect a raise in the bound electron polarizability due to lower constraints towards the air side than towards the bulk metal side. This could impact on the balance between free and bound electron terms in the total nonlinear surface susceptibility.

The bound electron spill out explored in scheme 3 is rather unusual. Matranga *et al.*[25] have shown that the SHG response of a metal surface with an adsorbed alkanethiol layer corresponds to an electron deficient interface as a consequence of the electronegativity of sulfur. As the on-top Au-S or Ag-S bond mostly involves the electrons from the 6s (resp. 5s) band[68], one may consider that these s electrons do not contribute to the free electron dielectric function anymore, because their polarizabilities are much reduced[21]. This is equivalently described by a decrease in the free electron density close to the interface, whereas the bound electron density is unchanged. Scheme 3 therefore takes into account the fact that the metal-S bonding creates a vacancy in free electrons leading to the situation symmetric to scheme 1, with the free electron density spilling less out than the bound electron one. The present case of sulfur covalently bound to gold and silver should not be confused with the situation of alkali adatoms or overlayers on top of transition metals, which lead to an enhancement of SHG as a consequence of the extreme polarizability of alkali metals[33]. We note that the situation should be different for an adsorbate-free metal surface, for which a true free electron spill out is expected. However, in this case, the method used in Ref. 1 would not be usable for the internal calibration of the evolution of the SFG intensities as a function of the visible wavelength.

We evaluate the free and bound electron polarizabilities from their respective dielectric functions, separated from the full tabulated dielectric function of silver and gold. Such a split is always somehow arbitrary, and alternate choices may alter the balance between free and bound contributions to the effective nonlinear susceptibility. The present modeling is self-consistent in the frame of classical mechanics and has proven its efficiency[7], but may still remain too coarse, for example when modeling the bound electrons as a continuous dipolium. In addition, the local Drude model, although widely used for such simulations, leads to a rapid growth of the dielectric function in the infrared, having a strong impact on the value of $a^f(\omega_1,\omega_2)$ under the long wavelength approximation. A more realistic model, either nonlocal[12,37,69], including a hydrodynamic pressure term[4,39,41], or even more sophisticated[43], and the limitation of the rise of $a^f$ [40], could help refine the definition of silver and gold properties and improve the accuracy of the modeled free electron response.

We see that the weight of the zzz contribution, with respect to the others, varies substantially between schemes 1, 2 and 3. In the scheme 3 for example, it even becomes smaller than the bulk and surface parallel terms above 500nm, and of the same order of magnitude below. This is not what is usually reported in the literature, where the surface perpendicular contribution is found bigger that the others[10,11,24] as in our schemes 1 and 2, in spite of unfavorable Fresnel coefficients (taking care of the definition of the $\chi^{(2)}$ tensor in terms of internal or external fields, and of the inclusion of some part of the bulk contribution into the surface terms[11]). However, as recalled in the introduction, most of the analyses published on the metal surfaces use SHG as a technique, which mostly differs from the SFG case by the absence of an infrared beam, characterized by a very high free electron and a very low bound electron contribution to the dielectric function. This dramatically impacts on the balance between the various terms, including through the modulation of Fresnel reflectivity, and prevents from a direct comparison of dispersion data between SFG and SHG. In addition, SHG studies often investigate bare metal surfaces, either in vacuum, air[70], water[24] or under electrochemical potential control[25,71]. As explained above, for such surfaces, the free electron term, and the zzz contribution, may indeed become dominant as a consequence of spill out.

### VI. Conclusion

In this paper, we applied a model for the calculation of the surface (parallel and perpendicular to the interface) and bulk contributions to the effective nonlinear susceptibility of silver and gold in an infrared-visible SFG experiment. The model explicitly takes into account the free and bound electron densities, and the dispersive behavior in the visible range only requires their respective dielectric functions as inputs. Our model is one of the few which takes into account the free and bound electrons together [31], and we aimed at doing so using a limited amount of unknown or free parameters. For that respect, we believe that it could be useful for a variety of applications. The perpendicular surface contribution requires additional hypotheses on the relative distributions of the free and bound electron density profiles across the interface. However, in the three cases considered, only their relative overlaps as a function of the depth across the interface and their bulk values matter, and not their actual shapes. The present study provides a general frame for the analysis of the intrinsic and effective nonlinear susceptibilities of gold and silver by second order nonlinear optics. An extension to the other metals is straightforward, and it should be interesting to consider the influence of the interband transitions on the nonlinear response in the cases of silver and platinum, for which they occur at higher and lower energy than the visible range, respectively. The three schemes describing the characteristic electron distributions make it adaptable to a variety of experimental cases. We believe that further improvements of this model will lead to a more sophisticated and accurate description of such systems.

As long as the bound electrons do not spill out of the bulk more than the free electrons, the latter, mainly through the surface perpendicular term, almost completely account for the total SFG response, even if the interband transitions modulate it through the SFG and visible dielectric functions. When the bound electrons are allowed to spill out of the free electron density, their contribution becomes more significant, but still minoritary except at high energy. Again the free electron dominates, this time through the surface parallel and bulk terms, and with a lower overall amplitude. This study proves that the actual shapes of the electron densities at the interface (or at least their overlaps), which are extremely difficult to experimentally measure on any given sample, should remain the key parameter in order to correctly analyze a gold surface by nonlinear optics.

In the next step, we plan to use this model in order to account for the experimental data detailed in Part I[1], by studying the influence of the balance between free and bound contributions, and a putative resonance with a surface state[67].

Appendix A: Quick review of the phenomenological description

Two light beams, modeled by plane waves at frequencies $\omega_1$ (visible, angle of incidence $\theta_1^1$) and $\omega_2$ (infrared, $\theta_2^1$) propagate in air (medium 1, $\varepsilon^{air}=\varepsilon^1=1$) and coincide at the metal surface (medium 2, dielectric function $\varepsilon^{metal} \equiv \varepsilon$, with $\varepsilon_i = \varepsilon(\omega_i) = (n_i)^2$ where n is the complex index of refraction of the metal), supposed isotropic. This hypothesis is reasonable when a polycrystalline surface is used in the experiments. Even if the (111) surface orientation is preferred in this case, macroscopic isotropy is recovered as a consequence of random orientation between the microscopic crystallites. The SFG beam ($\omega_3 = \omega_1 + \omega_2$) is detected in reflection at the phase matching angle $\theta_3$: $\omega_3 \sin\theta_3^1 = \omega_1 \sin\theta_1^1 + \omega_2 \sin\theta_2^1$ (superscripts refer to the medium, subscript to the beam). All beams are p-polarized. Across the interface, quantities $n_i \sin\theta_i$ remain constant in media 1 and 2: $\sin\theta_i^1 = n_i \sin\theta_i^2$. The thiol intermediate layer, considered non-absorbing and very thin as compared to all wavelengths, is blank for the propagation of light plane waves[44].

The incoming electric fields create in the bulk a nonlinear second-order dipolar polarization **P**(**r**,$\omega_3$), a quadrupolar polarization **Q**(**r**,$\omega_3$) and a magnetization **M**(**r**,$\omega_3$), each one quadratic in the local electric fields or their gradients through specific susceptibility tensors $\chi^{(2)}$ [18,72,73]. The dipolar polarization may be split into a "pure" dipolar term, proportional to the electric fields, and two terms involving also the field gradients and the magnetic fields[73]. These contributions are equivalent to a unique bulk polarization with[74]:

$$\mathbf{P}_B(\mathbf{r},\omega_3) = \mathbf{P}(\mathbf{r},\omega_3) - \frac{1}{2}\nabla \cdot \mathbf{Q}(\mathbf{r},\omega_3) + \frac{ic}{\omega_3}\nabla \times \mathbf{M}(\mathbf{r},\omega_3) \tag{A1}$$

For a cubic crystal, the pure dipolar term vanishes for symmetry reasons, and the bulk polarization involves only terms proportional to electric field gradients. Once **P**$_B$ is known all over the bulk, the total bulk contribution to the emission of SFG radiation has been calculated[2,21]. Additional surface terms arise from the gradients experienced by the electric fields when crossing the interface[9]. They are well described by an equivalent surface dipolar source[5], decomposed in terms parallel (called x/y below) and perpendicular (z) to the surface. This selvedge contribution is usually calculated in the long-wavelength approximation as a consequence of the small thickness of this interface slab. If translational symmetry is assumed in the surface plane, the local effective nonlinear polarization arises at the interface from the integration along z of the nonlocal contributions[75] and we may then define the surface polarization as[73]:

$$\mathbf{P}_S(\omega_3) = \int_I \mathbf{P}_B(z,\omega_3)dz \tag{A2}$$

where integration runs over the interface slab. It is parametrized by a surface nonlinear susceptibility tensor $\chi_S$ (we drop superscript (2) from now on for clarity), for which the formalism usually devoted to the emission of SFG radiation by molecular monolayers at interfaces applies[76]:

$$\mathbf{P}_S(\omega_3) = \chi_S(\omega_3; \omega_1, \omega_2) : \mathbf{E}(\omega_1)\mathbf{E}(\omega_2) \qquad (A3)$$

and the surface contribution to the SFG intensity becomes:

$$I_p(\omega_3) = \frac{8\pi^3(\omega_3)^2}{c^3 \cos^2\theta_3^1}\left|\chi_{S,ppp}^{eff}\right|^2 I_p(\omega_1) I_p(\omega_2) \text{ with } I_p(\omega_i) = \frac{c}{2\pi}\left|\mathbf{E}_p(\omega_i)\right|^2 \qquad (A4)$$

where the electric fields are evaluated in medium 1. In this equation, the effective surface nonlinear susceptibility tensor $\chi_S^{eff}$ encompasses the transmittivity and reflectivity conditions for the electric fields at the boundary, wrapped into Fresnel coefficients. We locate the sources of nonlinearity, and the electric fields which generate it, inside the metal. This is not the most common habit for SHG: in general, the incoming fields are evaluated inside the metal, but the nonlinear polarization outside[6,24,38,77]. However, our convention allows the best separation between intrinsic bulk and surface contributions in SHG[11]. Even if this choice has no impact on the effective and far field quantities, the definition of $\chi_S$ depends on this conventional choice[6,11,45].

We have, for an isotropic interface:

$$\chi_{S,ppp}^{eff} = F_{zzz}\chi_{S,zzz} - F_{xxz}\chi_{S,xxz} + F_{zxx}\chi_{S,zxx} - F_{xzx}\chi_{S,xzx} \qquad (A5a)$$

$$F_{zzz} = F_z(\omega_3)F_z(\omega_1)F_z(\omega_2)\sin\theta_3^1\sin\theta_1^1\sin\theta_2^1 \qquad (A5b)$$

$$F_{xxz} = F_x(\omega_3)F_x(\omega_1)F_z(\omega_2)\cos\theta_3^1\cos\theta_1^1\sin\theta_2^1 \qquad (A5c)$$

$$F_{zxx} = F_z(\omega_3)F_x(\omega_1)F_x(\omega_2)\sin\theta_3^1\cos\theta_1^1\cos\theta_2^1 \qquad (A5d)$$

$$F_{xzx} = F_x(\omega_3)F_z(\omega_1)F_x(\omega_2)\cos\theta_3^1\sin\theta_1^1\cos\theta_2^1 \qquad (A5e)$$

With the definitions above, the Fresnel factors for the surface terms have the form:

$$F_x(\omega_i) = \frac{2\cos\theta_i^2}{\cos\theta_i^2 + n_i\cos\theta_i^1} \text{ and } F_z(\omega_i) = \frac{2n_i\cos\theta_i^1}{\cos\theta_i^2 + n_i\cos\theta_i^1}\left(\frac{1}{n_i}\right)^2 \qquad (A6)$$

It can be shown[21,38] that the corresponding Fresnel factor for the bulk contribution is

$$F_{bulk} = F_z(\omega_3)F_z(\omega_1)F_z(\omega_2) \qquad (A7)$$

In a final step, the Rudnick-Stern dimensionless coefficients[10,25–27] allow to compare the models to a pure free electron gas response. Following the definitions of[38], we write:

$$\chi_{S,zzz} = \frac{1}{2n_0 e}\frac{\varepsilon_1 - 1}{4\pi}\frac{\varepsilon_2 - 1}{4\pi}\varepsilon_3 a(\omega_1, \omega_2) \qquad (A8a)$$

$$\chi_{S,xxz} = \frac{1}{2n_0 e}\frac{\varepsilon_1 - 1}{4\pi}\frac{\varepsilon_2 - 1}{4\pi}\frac{2\omega_1}{\omega_3}b(\omega_1, \omega_2) \qquad (A8b)$$

$$\chi_{S,xzx} = \frac{1}{2n_0 e}\frac{\varepsilon_1 - 1}{4\pi}\frac{\varepsilon_2 - 1}{4\pi}\frac{2\omega_2}{\omega_3}b(\omega_2, \omega_1) \qquad (A8c)$$

where $n_0$ is the bulk electron density; e the absolute value of the electron charge; indices 1,2,3 stand for visible, infrared and SFG, respectively. Parameters a and b refer to the incoming electric fields evaluated inside the metal, and creating the second order polarization outside, as is, to our knowledge, their universal convention. As we are interested in the absolute phases, the sign convention is also important. It depends on the orientation of the z-axis (towards or away from metal surface) and consequently on the integration in equation (A2). We adopt the convention usually used in SFG[21], with the z-axis pointing away from the interface, whereas Ref. 40, and several others[77] studying SHG, use the opposite. This choice accounts for the differences in the signs of a and b parameters and allows a coherent comparison of our results with the analysis of Ref. 1. Most available models show that the surface zxx term vanishes (although an effective zxx contribution originating in the bulk term is often included in the surface term[10,11]) because no gradients are experienced by the parallel components of the incoming electric fields[45]. Originally in SHG, parameter b equals -1 and is therefore dispersionless, contrary to a[4,25], whose amplitude may experiment resonant behaviors. However, all susceptibility components show dispersion in SFG (equation A8), and, as each term is modulated by the appropriate Fresnel coefficient, the overall effective terms all show a dependency on the incoming and generated wavelengths.

Appendix B: Coefficients for the bulk response

The D and Δ coefficients appearing in equation (3) have the following expressions:

$$D_i^{f/b} = \frac{1}{n_0^{f/b} e} \frac{\varepsilon_1^{f/b}-1}{4\pi} \frac{\varepsilon_2^{f/b}-1}{4\pi} d_i^{f/b} \quad \text{with} \quad d_i^{f/b} = \frac{\varepsilon_3^{f/b}-1}{\varepsilon_i^{f/b}-1} \frac{\omega_1 \omega_2}{\omega_i^2} \tag{B1}$$

$$\Delta_i^{f/b} = \frac{1}{n_0^{f/b} e} \frac{\varepsilon_1^{f/b}-1}{4\pi} \frac{\varepsilon_2^{f/b}-1}{4\pi} \delta_i^{f/b} \quad \text{with} \quad \delta_i^{f/b} = \frac{\omega_i}{\omega_3} - \frac{\varepsilon_3^{f/b}-1}{\varepsilon_i^{f/b}-1} \frac{\omega_3}{\omega_i} \tag{B2}$$

Appendix C: Calculation of the free and bound electron contributions to the SFG response

We follow the derivation of Ref. 36 and 38. The interface is defined as the region where the electron density varies from zero to its bulk value (and so does the dielectric function). Continuity is assumed for the perpendicular (z) and parallel (x) components of the displacement and electric fields, respectively. The local nonlinear polarization in this region, which sums up dipolar, quadrupolar and magnetic contributions, is therefore z-dependent. The full surface contribution to the nonlinear polarization at $\omega_3=\omega_{SFG}$ is obtained by integrating this z-dependent local value across the interface. Integration of the x-component is straightforward and leads to the result presented in the main text. In schemes 1, 2 and 3, the contributions from free and bound electrons are evaluated separately, then added up. The z-components of the local nonlinear polarizations induced by the free (resp. bound) electrons $P_{3z}^f(z)$ (resp. $P_{3z}^b$) are given by equation (16):

$$P_{B,z}^f(z) = \frac{\varepsilon_1 \varepsilon_2 \varepsilon_3}{e \varepsilon_3(z)} \left[ -n^f(z)\alpha_3^f \left( \frac{\alpha_1^f}{\varepsilon_1(z)} \frac{\partial}{\partial z} \frac{1}{\varepsilon_2(z)} + \frac{\alpha_2^f}{\varepsilon_2(z)} \frac{\partial}{\partial z} \frac{1}{\varepsilon_1(z)} \right) \right.$$
$$\left. + \alpha_1^f \alpha_2^f \frac{\partial}{\partial z} \left( n^f(z) \frac{1}{\varepsilon_1(z)} \frac{1}{\varepsilon_2(z)} \right) \right] E_{1z} E_{2z} \tag{C1}$$

where the quantities not depending explicitly on z are the bulk ones. Integration over the interface layer gives the surface nonlinear polarization:

$$P_{S,z}^f = \int_I P_{B,z}^f(z) dz \qquad (C2)$$

where integration may be extended to interval (-∞;+∞) for convenience.

Using $\varepsilon_i^{f/b}(z) = 1 + 4\pi\alpha_i^{f/b} n^{f/b}(z)$, the integrand is transformed into a sum of rational fractions of $n^f(z)$ and $n^b(z)$ only. The results of the main text may be deduced by use of the following procedures. The integrations are split between [$z_0$;+∞) and (-∞;$z_0$]. For scheme 1 (and conversely scheme 3 after swapping f and b superscripts), in the first half, $n^b(z)$ vanishes and $n^f(z)$ varies from $n_0^f$ to 0, thus:

$$\varepsilon_i(z) = 1 + 4\pi\alpha_i^f n^f(z) = 1 + \frac{n^f(z)}{n_0^f}\left[\varepsilon_i^f - 1\right] = 1 + u\left(\varepsilon_i^f - 1\right). \qquad (C3)$$

The integration variable is changed from z to $u = n^f(z)/n_0^f$, which varies from 1 to 0. In the second half, $n^f(z)$ remains equal to $n_0^f$ while $n^b(z)$ varies from $n_0^b$ to 0. Integration in this case runs over variable $u = n^b(z)/n_0^b$, varying from 1 to 0, using:

$$\varepsilon_i(z) = \varepsilon_i^f + 4\pi\alpha_i^b n^b(z) = \varepsilon_i^f\left[1 + \frac{n^b(z)}{n_0^b}\frac{(\varepsilon_i^b - 1)}{\varepsilon_i^f}\right] = \varepsilon_i^f\left[1 + u\frac{(\varepsilon_i^b - 1)}{\varepsilon_i^f}\right]. \qquad (C4)$$

In scheme 2, both electron densities vary together and $\varepsilon_i(z) = 1 + 4\pi\alpha_i^f n^f(z) + 4\pi\alpha_i^b n^b(z)$.

Integration may be performed by using the constant $R = \frac{n^b(z)}{n^f(z)} = \frac{n_0^b}{n_0^f}$, which implies

$$\varepsilon_i(z) = 1 + 4\pi\alpha_i^f n^f(z) + 4\pi\alpha_i^b n^b(z) = 1 + 4\pi n^f(z)\left[\alpha_i^f + R\alpha_i^b\right] = 1 + (\varepsilon_i - 1)\frac{n^f(z)}{n_0^f}. \qquad (C5)$$

One may then choose $u = n^f(z)/n_0^f$ as the integration variable, as for scheme 1. The calculations involve integrals of the form:

$$J_0(q_1, q_2, q_3) = \int_0^1 \frac{dx}{(1+q_1 x)(1+q_2 x)(1+q_3 x)} \equiv J_0, \text{ with } q_1 \neq q_2 \neq q_3 \qquad (C6)$$

$$I_0(q_1, q_2; q_3) = \int_0^1 \frac{dx}{(1+q_1 x)(1+q_2 x)(1+q_3 x)^2} \equiv I_0(q_3) \qquad (C7)$$

$$I_1(q_1, q_2; q_3) = \int_0^1 \frac{x \, dx}{(1+q_1 x)(1+q_2 x)(1+q_3 x)^2} \equiv I_1(q_3) \qquad (C8)$$

A direct calculation is possible after partial fraction decomposition. As an example,

$$J_0 = \frac{q_1}{(q_1-q_2)(q_1-q_3)}\ln(1+q_1) + \text{c.p.} = \sum_{\substack{\{i,j,k\}= \\ \text{c.p. of} \\ \{1,2,3\}}} \frac{q_i}{(q_i-q_j)(q_i-q_k)}\ln(1+q_i) \qquad (C9)$$

where c.p. denotes circular permutation of $q_1$, $q_2$, $q_3$.

For the calculation of formula (C1) above, in order for example to recover symmetric expressions between 1, 2 and 3 indices, the reader may find the following relations useful:

$$q_i \, I_1(q_i) + I_0(q_i) = J_0 \qquad (C10)$$

$$I_0(q_1) + I_0(q_2) + I_0(q_3) = 2J_0 + \frac{1}{(1+q_1)(1+q_2)(1+q_3)} \qquad (C11)$$

$$I_1(q_1) + I_1(q_2) + I_1(q_3) = \frac{1}{(1+q_1)(1+q_2)(1+q_3)} - \sum_{\substack{\{i,j,k\}= \\ \text{c.p. of} \\ \{1,2,3\}}}^{3} \frac{\ln(1+q_i)}{(q_i - q_j)(q_i - q_k)} \qquad (C12)$$

$$q_1 I_0(q_1) + q_2 I_0(q_2) + q_3 I_0(q_3) = 1 - \frac{1}{(1+q_1)(1+q_2)(1+q_3)} \qquad (C13)$$

$$q_1 I_1(q_1) + q_2 I_1(q_2) + q_3 I_1(q_3) = J_0 - \frac{1}{(1+q_1)(1+q_2)(1+q_3)} \qquad (C14)$$

Knowing the expression of the surface nonlinear polarization (C2) leads to the surface $\chi^{(2)}$ components (equation (A3)), which may be expressed as functions of $a^{f/b}(\omega_1,\omega_2)$ and $b^{f/b}(\omega_1,\omega_2)$ through equations (7), (9) and (11), and plugged into equation (A5) to calculate the effective susceptibilities.